\documentclass[aip,apl,twocolumn,superscriptaddress,reprint]{revtex4-1}

\usepackage{graphicx}
\usepackage{amsmath}
\usepackage{dcolumn}

\usepackage{color}

\newcommand{\Eq}[1]{Eq.~\eqref{#1}}

\begin{document}
	\title{Period multiplication in a parametrically driven superconducting resonator}
	
	\author{Ida-Maria \surname{Svensson}}
	\email[]{ida-maria.svensson@chalmers.se}
	\author{Andreas \surname{Bengtsson}}
	%\author{Jonathan \surname{Burnett}}
	\author{Jonas \surname{Bylander}}
	\author{Vitaly \surname{Shumeiko}}
	\author{Per \surname{Delsing}}
	\email[]{per.delsing@chalmers.se}
	\affiliation{Microtechnology and Nanoscience, Chalmers University of Technology, SE-41296 G\"{o}teborg, Sweden}
	\date{\today}
	
\begin{abstract}
We report on the experimental observation of period multiplication in parametrically driven tunable superconducting resonators. We modulate the magnetic flux through a superconducting quantum interference device, attached to a quarter-wavelength resonator, with frequencies $n\omega$ close to multiples, $n=2,\,3,\,4,\,5$, of the resonator fundamental mode  and observe intense output radiation at $\omega$. The output field manifests $n$-fold degeneracy with respect to the phase, the $n$ states are phase shifted by $2\pi/n$ with respect to each other. Our demonstration verifies the theoretical prediction by Guo et al. \cite{Guo2013}, and paves the way for engineering complex macroscopic quantum cat states with microwave 
photons.
\end{abstract}
	
\maketitle	
	
%%%%%%%%%%%%%%%%
%\section{Introduction}
The technology of circuit quantum electrodynamics \cite{Schoelkopf2008,Gu2017} offers an excellent platform for observation and exploration of parametric oscillation phenomena in the quantum domain. By connecting Josephson elements to superconducting resonators, one is able to induce nonlinearity of the electromagnetic field and realise temporal high frequency control of the resonator parameters \cite{Wallquist2006,Sandberg2008,PalaciosLaloy2008}. This in combination with high quality factors of the superconducting resonators makes the parametric oscillation regime, above the instability threshold, easily accessible with relatively small modulation intensities. Furthermore, low temperatures in the range of 10 mK allows to investigate the quantum properties of the oscillator states. Using this technique, both the degenerate (pumping at twice a resonator mode frequency) and non-degenerate (pumping at the sum of two resonator mode frequencies) parametric oscillations, have been experimentally investigated \cite{Wilson2010,Bengtsson2018paper}.

An inherent property of parametric oscillations is phase degeneracy of the oscillator states. The non-degenerate oscillator has a continuous phase degeneracy \cite{Bengtsson2018paper,Wustmann2017,Sun2016}, while the degenerate oscillator exhibits a discrete, two-fold degeneracy, which is manifested by two correlated $\pi$-shifted steady states \cite{Wilson2010,Wustmann2013}. In the quantum regime these states form coherent superpositions of optical coherent states, cat states \cite{Vlastakis2013,Puri2017}, which can be used as building blocks for a photonic quantum processor \cite{Mirrahimi2014}. 

Quantum properties of the degenerate parametric oscillations motivate a great interest in finding ways to engineer more complex multiply degenerate oscillator states. The period multiplication phenomenon in nonlinear oscillators \cite{Hayashi,JordanSmith} offers an attractive approach to the problem. In recent papers, an experimental demonstration of the period tripling in a superconducting resonator was reported \cite{Svensson2017PRB}, and quantum properties of the emerging three-fold degenerate state were theoretically investigated \cite{Zhang2017}. In that experiment, the self-sustained oscillations of the resonator mode were excited by injecting an external signal with a frequency close to three times the fundamental mode frequency; the excitation mechanism involves, first, an excitation of a higher resonator mode with frequency close to the driving frequency, and then a parametric down-conversion of this excited mode field. Another method for period multiplication was theoretically proposed by Guo \textit{et al.} \cite{Guo2013}, based on a modulation of the oscillator nonlinearity rather than the frequency. If the oscillator potential contains a nonlinear term, $\propto \epsilon q^{n+1}$, then a modulation of the coefficient $\epsilon$, with a frequency  $n\omega$, will produce $n$-fold degenerate self-sustained oscillations with frequency $\omega$.

In this Letter we report an experimental demonstration of period multiplication in a parametrically driven superconducting resonator, and the emergence of multiply degenerate, self-sustained oscillations. To do this we use a frequency-tunable, quarter-wavelength coplanar waveguide microwave resonator \cite{Sandberg2008,Wallquist2006,PalaciosLaloy2008}. The resonator is capacitively coupled to a transmission line at one end, and grounded at the other end via a superconducting quantum interference device (SQUID). Due to the nonlinearity of the SQUID, the frequency spectrum of the resonator is non-equidistant, the lowest mode having a frequency close to $5\,$GHz. A similar device has been used earlier to demonstrate degenerate parametric oscillations  \cite{Wilson2010} by modulating the SQUID inductance at two times the frequency of the resonator mode.  Note that the modulation of the SQUID inductance not only affects the quadratic term, \textit{i.e.} the resonator frequency, but also all the higher order nonlinear terms of the Taylor expansion of the differential inductance,
\begin{equation}
L_{SQ}(\phi)=\frac{L_{SQ,0}}{\cos (f/2)\cos \phi   - ( E_-/E_+) \,\sin (f/2)\sin \phi }.
\end{equation}
Here $f(t)= 2\pi \Phi(t)/\Phi_0$ is a normalized applied magnetic flux,  $\Phi_0=h/2e$ is the flux quantum, $E_\pm=(E_{J1}\pm E_{J2})/2$,  are combinations of Josephson energies of the two SQUID junctions, and $L_{SQ,0}=(\hbar/2e)^2(1/2E_+)$ is the SQUID inductance at zero flux. 
Thus, by choosing an appropriate modulation frequency, one can selectively address any of the nonlinear terms and hence implement the period multiplication proposed in Ref.~[\onlinecite{Guo2013}].   

%%%%%%%%%%%%%%%%%%%%
	
The samples used are fabricated using standard micro-fabrication techniques. The circuits are etched in a thin niobium film on c-plane sapphire substrates. The SQUIDs are made of aluminium, and fabricated using two-angle evaporation. 	
\begin{figure}
		\centering
		\includegraphics{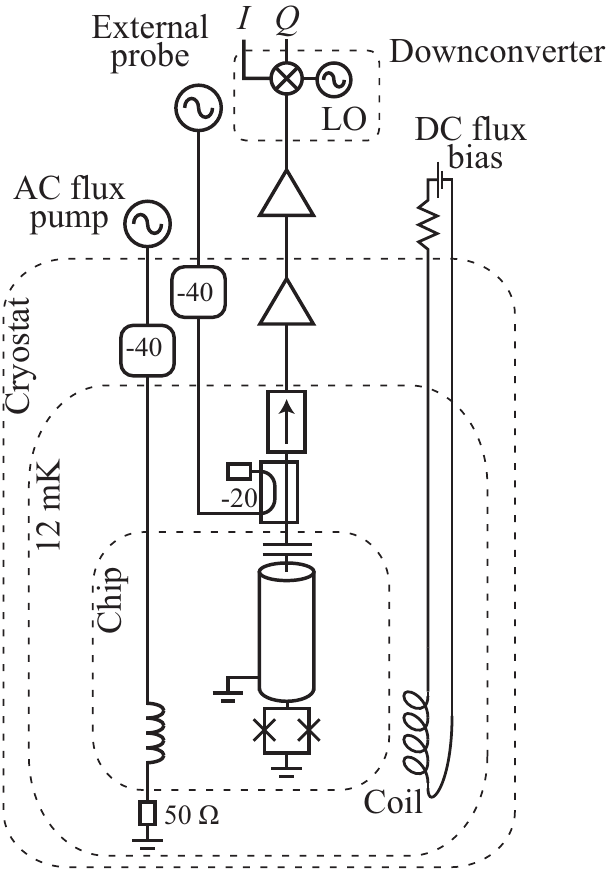}
		\caption{\label{fig:setup}Schematic of the measurement setup. The resonator can be  excited via an external probe or via the flux pump line. The chip is mounted in a sample box at the mixing chamber stage of a dilution cryostat with a base temperature of $12\,$mK. Measurements are done using heterodyne detection.}
	\end{figure}
The measurements are performed at low temperatures, $\hbar\omega\gg k_BT$, in a dilution cryostat, where the base temperature is around $12\,$mK. Our measurement setup is sketched in Fig.~\ref{fig:setup}. We use two input lines, one for an external probe signal and one for flux modulation. Both are attenuated to reduce thermal noise. A static magnetic field can be applied using a superconducting coil, mounted close to the sample. 
	
The sample is measured using a reflection setup. A directional coupler is employed to route the signal, and isolators protect the resonator from amplifier noise. 		
The output signal is amplified by a cryogenic amplifier and then further boosted by a room temperature amplifier. The signal undergoes heterodyne down-conversion and the in- and out-of-phase quadrature voltages $I(t)$ and $Q(t)$ are then sampled in a digitizer.
	
%	\section{Results}
\begin{figure}
	\centering
	\includegraphics{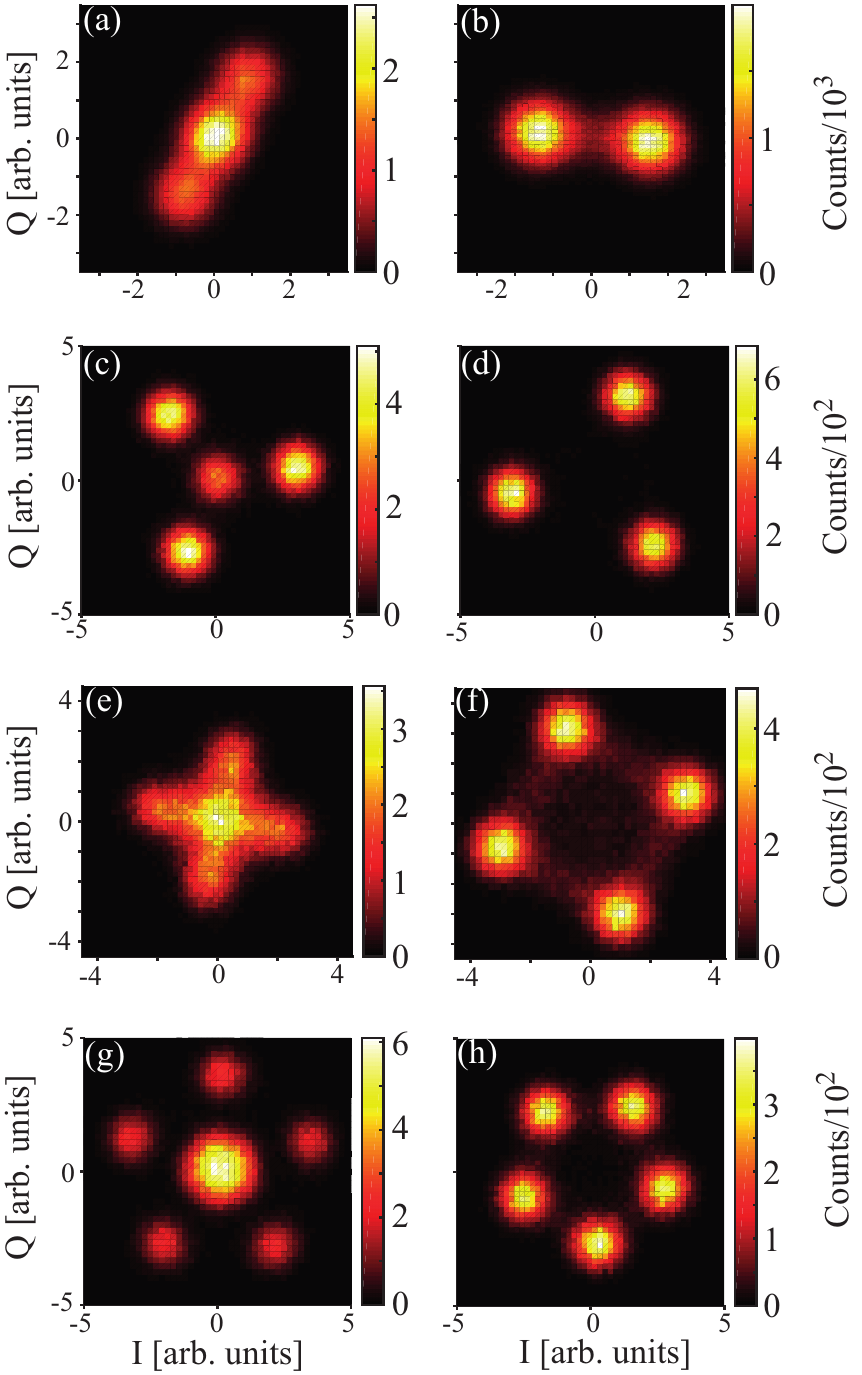}
	\caption{\label{fig:SOnw}Subharmonic oscillations generated by applying a microwave signal to the flux pump line at multiples $n$ of the fundamental mode frequency. (a,b) $n=2$; (c,d) $n=3$; (e,f) $n=4$; (g,h) $n=5$.}
\end{figure}
Our measurement frequency is placed close to the fundamental mode, $\omega = \omega_1+\delta$, and a microwave tone is applied to the flux pump line at a frequency $n\omega$, $n$ = 2, 3, 4, 5.
For a small pump intensity we only detect noise, while when the pump intensity exceeds a certain threshold, a strong output signal emerges indicating excitation of the parametric oscillator.  

We analyze the output by sampling the quadratures and produce the histograms shown in Fig.~\ref{fig:SOnw}. The bright spots in the histograms correspond to steady states of the oscillator; they clearly indicate a discrete degeneracy of the oscillator states. The plots in the right column of Fig.~\ref{fig:SOnw} show spots that are symmetrically displaced from the origin, \textit{i.e.} have equal intensities, $P = I^2 + Q^2$,  and are phase shifted by $2\pi/n$ with respect to each other. Overall orientations of the multiplets are defined by the phase of the pump signal (as will be shown below), which we do not control in this experiment. Hence, the orientations of different histograms are random. 

The histograms in the right column of Fig.~\ref{fig:SOnw} were measured in the well-established oscillator regime with pump amplitude well above the instability threshold. Those to the left were measured closer to the threshold, and here we also see a central spot representing the oscillator ground state. This is a multistability region, which is typical for parametric oscillations, as explained theoretically \cite{Wustmann2013,Svensson2017PRB} and observed experimentally for the degenerate parametric oscillator \cite{Wilson2010} and for the period-tripling oscillator \cite{Svensson2017PRB}. The multistability is explained by the fact that the oscillator ground state maintains stability and may coexist with the excited states. In histograms (a) and (e) the excited states are not well resolved. This is due to large critical fluctuations in the parameter region, where the coherent oscillations emerge having a small relative intensity. Transitions between the different states occur continuously. In histogram (f) these transitions are observed as faint lines between the spots. This is an averaging effect of stochastic switching occurring during sampling.

To explain our observations we analyze the multimode Hamiltonian of the resonator \cite{Wustmann2013}, 
\begin{eqnarray}\label{H}
H &=& \sum_n \hbar\omega_n \hat{a}^\dag_n \hat{a}_n + V(\hat{\phi}, t)\,, 
\end{eqnarray}
where $\omega_n$ is the frequency of the $n$-th eigenmode, $\hat{a}_n$ and $\hat{a}^\dag_n$ are the mode annihilation and creation operators, and the sum goes over all resonator modes. The variable 
$\hat{\phi}(t)$ refers to a dynamic deviation from the static value, $\tan\phi_0= -(E_-/E_+)\tan(F/2)$, of the phase at the SQUID-terminated edge of the resonator, $\hat{\phi}(x=d,t) = \phi_0 + \hat{\phi}(t)$. Here, $F=2\pi\Phi_{dc}/\Phi_0$ indicates a normalized static magnetic flux, and $d$ denotes the resonator length. An expansion of $\hat{\phi}$ over the cavity modes reads, 
\begin{eqnarray}\label{phi}
\hat{\phi} =  \sum_n \beta_n \,(\hat{a}_n+\hat{a}^\dag_n), \quad 
\beta_n = \gamma\,\sqrt{8\pi Z_0 k_nd \over R_K} \, , 
\end{eqnarray}
where $Z_0= \sqrt{L_0/ C_0}, \;  R_K = {h/ e^2} $, and $k_n= \omega_n/v$ is the mode eigenvector defined by the spectral equation\cite{Wustmann2013}, $k_nd\tan(k_nd) = 1/\gamma$, where $\gamma = E_{L,cav}/(2E_+\cos(F/2)) \ll1$ is the participation ratio of the SQUID inductance versus the cavity inductance. Here, $L_0$ and $C_0$ are the inductance and capacitance per unit length of the coplanar waveguide transmission line, respectively. Together they define the phase velocity in the waveguide resonator, $v=1/\sqrt{L_0C_0}$. The potential energy $V(\hat{\phi},t)$ in Eq.~\eqref{H} represents the nonlinear part of the inductive energy of the SQUID and has the form for an asymmetric SQUID,
\begin{eqnarray}\label{V}
&& V(\hat{\phi}, t)  \approx  -E_+\cos(F/2)\left(\cos \hat{\phi} + {\hat{\phi}^2\over 2} \right) \nonumber\\
&+& \delta f(t) \left[   E_+\sin(F/2)\cos\hat{\phi} +  {E_-\over \cos(F/2)} \sin\hat{\phi}\right]\,,
\end{eqnarray}
where $\delta f(t)=2\pi\Phi_{ac}(t)/\Phi_0$ describes the pump, \textit{i.e.} a small-amplitude temporal modulation of the applied magnetic flux, which we here consider in the linear approximation. Other adopted approximations in \Eq{V} include the assumption of a small SQUID asymmetry, $E_-\ll E_+ \ll 1$, and that the effect of the SQUID capacitance is small. 
 
It is appropriate to make the following comments to \Eq{V}: (i) the symmetric part of the potential, $\propto E_+$, contains only even powers of $\hat{\phi}$, which implies that only even subharmonics can be excited by the flux pump if the SQUID is perfectly symmetric. (ii) The asymmetric part, $\propto E_-$, is responsible for the excitation of odd subharmonics; this part also contains a linear term, $\propto \hat{\phi}$, which implies that the nonstationary magnetic flux directly generates a field inside the resonator, \textit{i.e.} acts as an effective current drive, in addition to the parametric pump. This field is down-converted via the mechanism discussed in detail in [\onlinecite{Svensson2017PRB}] and contributes to excitation of odd subharmonics: for odd subharmonics, a higher resonator mode lies close to the driving frequency, and this mode is efficiently excited by the drive and acts as a pump exciting the subharmonics. (iii) There is a possibility of a parasitic crosstalk between the flux pump line and the resonator that may effectively give a current drive of the resonator at the pump frequency. Therefore, it is possible to get both odd and even subharmonics, even if the SQUID is completely symmetric. Both the crosstalk and the asymmetry of the SQUID are difficult to estimate and this makes it hard to make a quantitative comparison to the theory.

Suppose the magnetic flux is modulated with a frequency $n\omega = n(\omega_1 +\delta)$ and $\delta f(t) = \delta f_0\cos n\omega t$. Then we assume, focusing on the resonant response of the fundamental resonator mode $n=1$,  the field in the cavity to be a superposition of two harmonics,
\begin{eqnarray}\label{phi_ansatz}
\hat{\phi}(t) =   \beta _1 \hat{a}_1(t)e^{-i\omega t} + \beta_n \hat{a}_n(t)e^{-in\omega t}  + {\rm H.c.} 
\end{eqnarray}
where $\hat{a}_n$ refers to a response of a mode with a frequency close to the pump frequency. A quasiclassical approximation is relevant for large-amplitude subharmonic oscillations far from the threshold. Following the method of Refs.~[\onlinecite{Wustmann2013,Svensson2017PRB}] we derive a shortened dynamical equation for a slowly varying quasiclassical amplitude 
$a_1(t)$ in the rotating wave approximation. This equation has the following universal form for all integers $n$,
\begin{eqnarray}\label{EOM}
i\dot a_1 + (\delta + i\Gamma_1 + \alpha |a_1|^2) a_1 + \epsilon_n (a_n^\ast)^{n-1} = 0 \,. 
\end{eqnarray}
The coefficients,  $\alpha =(E_+/\hbar)\cos(F/2) \beta_1^4$ and 
$\epsilon_2 =(E_+/\hbar)\cos(F/2) \beta_1^2\delta f_0$, are familiar from the study of degenerate parametric resonance\cite{Wustmann2013}. The  pump coefficient for period quadrupling, $\epsilon_4$, is expressed through $\epsilon_2$, 
$\epsilon_4 = -\epsilon_2 (\beta_1^2/2)$. Similar scaling holds for higher even-order coefficients, $\epsilon_{2k} \propto \epsilon_2 \beta_1^{2k-2}$. The pump coefficient for period tripling consists of two contributions, 
\begin{eqnarray}\label{epsilon3}
\epsilon_3 = { E_-  \over 2\hbar}{\beta_1^3\delta f_0 \over \cos(F/2)} +  {E_+\over \hbar}\cos(F/2)\beta_1^2\beta_3 a_3 \,. 
\end{eqnarray}
The first term is the direct effect of the flux pump and it only exists for an asymmetric SQUID. The second term results from the secondary, current-driving effect of the higher mode; this term exists for both symmetric and asymmetric SQUIDs. All higher odd-order coefficients have similar structure with the scaling factor $\beta_1^{2k-2}$. 

The stationary solutions of \Eq{EOM} for flux pumping at the $n$-th multiple of the fundamental resonator mode have the form $a_{1,n} = r_ne^{i\theta_n}$, with 
	$\sin(n\theta_n- {\rm arg}\,\epsilon_n) = \Gamma_1/|\epsilon|r_n^{n-2}$.
Such solutions have an $n$-fold degeneracy, as seen in Fig.~\ref{fig:SOnw},  with phases $\theta_n = \theta_{n,0} + 2\pi m/n$ $(m=0\ldots n-1)$. The reference phase $\theta_{n,0}$ depends on the phase of the pump and the parameters of the working point.

The outlined calculations justify the relevance of the model \cite{Guo2013} for parametrically driven superconducting resonators. They provide explicit equations for the model coefficients for this setup and qualitatively explain our experimental observations.

\begin{figure}
	\centering
	\includegraphics{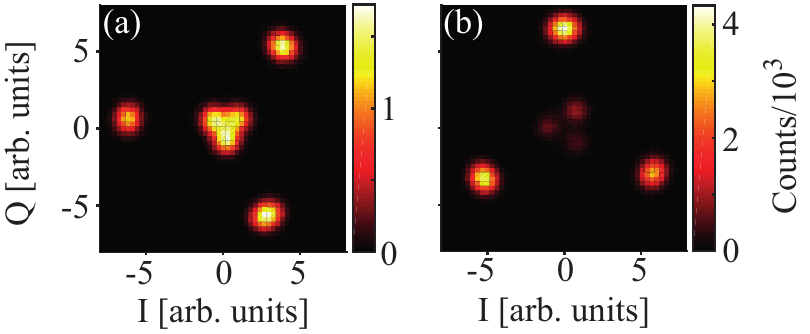}
	\caption{\label{fig:sixblobs}Histograms of period-tripling oscillations at larger output signal levels, revealing an additional triplet of excited states with amplitude and orientation different from the main triplet. The difference between panel (a) and (b) is that the later is measured at $1\,$dB higher pump power.}
\end{figure}

%%%%%%%%%%%%%%%%%%%%%%%%%
%
The nonlinear parametrically driven oscillator has indeed a very rich behavior. Further insight into the properties of period multiplication can be gained by exploring regimes beyond the observed steady-state multiplets. In Fig.~\ref{fig:sixblobs} we present a histogram of the period-tripling oscillations measured at larger red detuning, where the oscillation intensity is substantially stronger than the one in the histograms in Fig.~\ref{fig:SOnw}. Here we detect coexistence of two triplet states, with different amplitudes and different orientations, presumably corresponding to a higher order nonlinearity.

\begin{figure}
	\centering
	\includegraphics{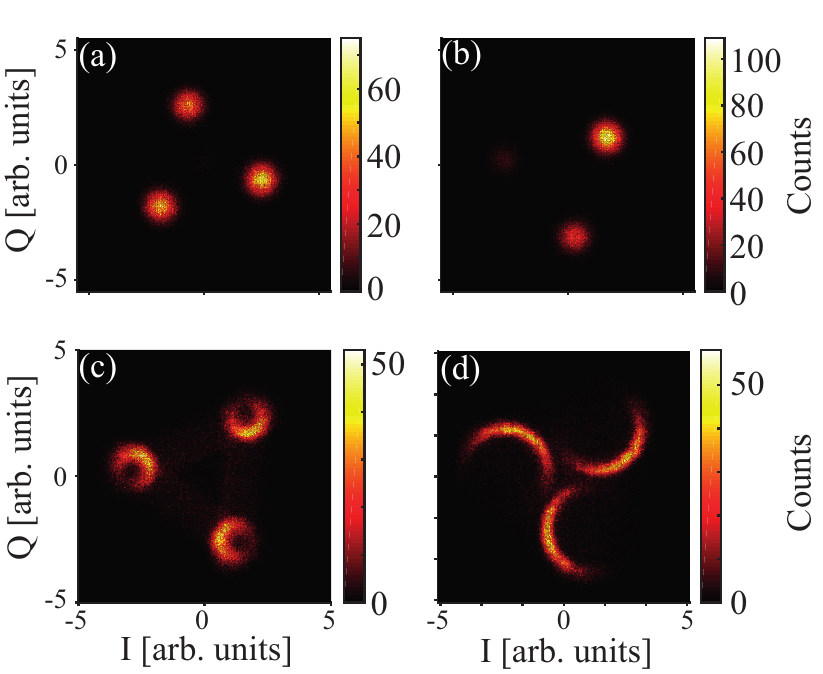}
	\caption{\label{fig:helicopter} Histograms of output oscillations affected by a probe signal. (a)-(b) Phase locking effect of an on-resonance probe, $20\,$dB stronger in (b) than in (a). (c)-(d) Effect of a $1\,$Hz detuned probe signal, $10\,$dB stronger in (d) than in (c). }
\end{figure}

Injection of a weak probe  signal into the resonator is known to produce a phase-locking effect, observed for example in the degenerate parametric oscillator  \cite{Lin2014}. There, the effect was manifested by a gradual disappearance of one of the doublet states, and explained by a symmetry breakdown under on-resonance injection. Similarly, in the period tripling regime we observe a gradual disappearance of the triplet components when an on-resonance probe signal is applied, see Fig.~\ref{fig:helicopter}(a) and (b). 

Furthermore, injection of a probe signal slightly detuned, by 1 Hz, produces a completely different, dramatic effect on the period tripling oscillations, see Fig.~\ref{fig:helicopter}(c) and (d). Instead of breaking the symmetry of the overall triplet pattern, the individual round spots deform into crescents with a size that increases with increasing probe intensity. Such a behavior resembles the phase locking effect observed in non-degenerate  parametric oscillators \cite{Bengtsson2018paper,Sun2016}, where the oscillator state possesses continuous phase degeneracy. 
This resemblance leads us to interpret our observation as the result of a deformation of each stable stationary state of the triplet into a stable cycle with simultaneous phase locking. A further theoretical investigation is required to give quantitative explanations to these observations.	
	
%	\section{Conclusion}
In conclusion, we observed period-multiplication phenomena in parametrically driven superconducting resonators. 
We observed robust output radiation at a frequency close to the fundamental resonator mode, with $n=2,3,4,5$ evenly shifted phase components under an applied pump signal with frequencies $n$ times the detection frequency. Our qualitative analysis of the resonator dynamics agrees with the observations, and corroborates the model proposed in [\onlinecite{Guo2013}]. Our observations put a firm ground for further exploration of quantum aspects of the period-multiplication phenomena, and the possibility of engineering complex photonic cat states with potential applications to information technologies.	

%\section{Acknowledgments}
We gratefully acknowledge financial support from the Wallenberg Foundation, the Swedish research council, and the European Research Council. J.B. acknowledges partial support by the EU under REA grant agreement no. CIG-618353.


\begin{thebibliography}{10}
	
	\bibitem{Guo2013}
	L.~Guo, M.~Marthaler, and G.~Sch\"on.
	\newblock Phase space crystals: A new way to create a quasienergy band
	structure.
	\newblock {\em Physical Review Letters}, 111:205303, 11 2013.
	
	\bibitem{Schoelkopf2008}
	R.~J. Schoelkopf and S.~M. Girvin.
	\newblock Wiring up quantum systems.
	\newblock {\em Nature}, 451(7179), 2008.
	
	\bibitem{Gu2017}
	Xiu Gu, Anton~Frisk Kockum, Adam Miranowicz, Yu-xi Liu, and Franco Nori.
	\newblock Microwave photonics with superconducting quantum circuits.
	\newblock {\em Physics Reports}, 2017.
	
	\bibitem{Wallquist2006}
	M.~Wallquist, V.~S. Shumeiko, and G.~Wendin.
	\newblock Selective coupling of superconducting charge qubits mediated by a
	tunable stripline cavity.
	\newblock {\em Physical Review B}, 74(22):224506, 12 2006.
	
	\bibitem{Sandberg2008}
	M.~Sandberg, C.~M. Wilson, F.~Persson, T.~Bauch, G.~Johansson, V.~Schumeiko,
	T.~Duty, and P.~Delsing.
	\newblock Tuning the field in a microwave resonator faster than the photon
	lifetime.
	\newblock {\em Applied Physics Letters}, 92:203501, 5 2008.
	
	\bibitem{PalaciosLaloy2008}
	A.~Palacios-Laloy, F.~Nguyen, F.~Mallet, P.~Bertet, D.~Vion, and D.~Esteve.
	\newblock Tunable resonators for quantum circuits.
	\newblock {\em Journal of Low Temperature Physics}, 151(3):1034, 2008.
	
	\bibitem{Wilson2010}
	C.~M. Wilson, T.~Duty, M.~Sandberg, F.~Persson, V.~Shumeiko, and P.~Delsing.
	\newblock Photon generation in an electromagnetic cavity with a time-dependent
	boundary.
	\newblock {\em Physical Review Letters}, 105(23):233907, 12 2010.
	
	\bibitem{Bengtsson2018paper}
	Andreas Bengtsson, Philip Krantz, Micha\"el Simoen, Ida-Maria Svensson, Ben~H.
	Schneider, Vitaly Shumeiko, Per Delsing, and Jonas Bylander.
	\newblock Nondegenerate parametric oscillations in a tunable superconducting
	resonator.
	\newblock arXiv:1801.04566.
	
	\bibitem{Wustmann2017}
	Waltraut Wustmann and Vitaly Shumeiko.
	\newblock Nondegenerate parametric resonance in a tunable superconducting
	cavity.
	\newblock {\em Physical Review Applied}, 8(2):024018, 08 2017.
	
	\bibitem{Sun2016}
	Fengpei Sun, Xiaoshi Dong, J~Zou, Mark~I Dykman, and Ho~Bun Chan.
	\newblock Correlated anomalous phase diffusion of coupled phononic modes in a
	sideband-driven resonator.
	\newblock {\em Nature communications}, 7:12694, 2016.
	
	\bibitem{Wustmann2013}
	Waltraut Wustmann and Vitaly Shumeiko.
	\newblock Parametric resonance in tunable superconducting cavities.
	\newblock {\em Physical Review B}, 87(18):184501, 5 2013.
	
	\bibitem{Vlastakis2013}
	Brian Vlastakis, Gerhard Kirchmair, Zaki Leghtas, Simon~E. Nigg, Luigi Frunzio,
	S.~M. Girvin, Mazyar Mirrahimi, M.~H. Devoret, and R.~J. Schoelkopf.
	\newblock Deterministically encoding quantum information using 100-photon
	schr{\"o}dinger cat states.
	\newblock {\em Science}, 342(6158):607, 2013.
	
	\bibitem{Puri2017}
	Shruti Puri, Samuel Boutin, and Alexandre Blais.
	\newblock Engineering the quantum states of light in a kerr-nonlinear resonator
	by two-photon driving.
	\newblock {\em NPJ Quantum Information}, 3(18), 2017.
	
	\bibitem{Mirrahimi2014}
	Mazyar Mirrahimi, Zaki Leghtas, Victor~V Albert, Steven Touzard, Robert~J
	Schoelkopf, Liang Jiang, and Michel~H Devoret.
	\newblock Dynamically protected cat-qubits: a new paradigm for universal
	quantum computation.
	\newblock {\em New Journal of Physics}, 16(4):045014, 2014.
	
	\bibitem{Hayashi}
	C.~Hayashi.
	\newblock {\em Nonlinear Oscillations in Physical Systems}.
	\newblock Princeton, 1985.
	
	\bibitem{JordanSmith}
	D.~W. Jordan and P.~Smith.
	\newblock {\em Nonlinear Ordinary Differential Equations}.
	\newblock Oxford, 2007.
	
	\bibitem{Svensson2017PRB}
	Ida-Maria Svensson, Andreas Bengtsson, Philip Krantz, Jonas Bylander, Vitaly
	Shumeiko, and Per Delsing.
	\newblock Period-tripling subharmonic oscillations in a driven superconducting
	resonator.
	\newblock {\em Physical Review B}, 96(17):174503, 11 2017.
	
	\bibitem{Zhang2017}
	Yaxing Zhang, J.~Gosner, S.~M. Girvin, J.~Ankerhold, and M.~I. Dykman.
	\newblock Time-translation-symmetry breaking in a driven oscillator: From the
	quantum coherent to the incoherent regime.
	\newblock {\em Physical Review A}, 96(5):052124, Nov 2017.
	
	\bibitem{Lin2014}
	Z.~R. Lin, K.~Inomata, K.~Koshino, W.~D. Oliver, Y.~Nakamura, J.~S. Tsai, and
	T.~Yamamoto.
	\newblock Josephson parametric phase-locked oscillator and its application to
	dispersive readout of superconducting qubits.
	\newblock {\em Nature Communications}, 5:4480, 7 2014.
	
\end{thebibliography}
\end{document}